\documentclass[aps,pra,superscriptaddress,floatfix,10pt, twocolumn]{revtex4}
\usepackage{bbm}
\usepackage{amsfonts}
\usepackage[dvips]{graphicx}
\usepackage{amsmath,amsfonts,amssymb,graphics,graphicx,epsfig,color,times,bbm}

\usepackage{lipsum}
\usepackage{graphicx}
\usepackage{epstopdf}
\usepackage{subfigure}
\if CLASSOPTIONcompsoc
\usepackage[caption=false, font=normalsize, labelfont=sf, textfont=sf]{subfig}
\else
\usepackage[caption=false, font=footnotesize]{subfig}
\fi
\DeclareMathOperator{\Tr}{Tr}

\begin{document}

\title{Experimental demonstration of coherence flow in $\mathcal{PT}$- and anti-$\mathcal{PT}$-symmetric systems}

\author{Yu-Liang Fang}
\thanks{These authors contributed equally}
\affiliation{Quantum Information Research Center, Shangrao Normal University, Shangrao 334000, China}

\author{Jun-Long Zhao}
\thanks{These authors contributed equally}
\affiliation{Quantum Information Research Center, Shangrao Normal University, Shangrao 334000, China}

\author{Yu Zhang}
\thanks{smutnauq@smail.nju.edu.cn}
\affiliation{Quantum Information Research Center, Shangrao Normal University, Shangrao 334000, China}
\affiliation{School of Physics, Nanjing University, Nanjing 210093, China}

\author{Dong-Xu Chen}
\affiliation{Quantum Information Research Center, Shangrao Normal University, Shangrao 334000, China}

\author{Qi-Cheng Wu}
\thanks{wqc@sru.edu.cn}
\affiliation{Quantum Information Research Center, Shangrao Normal University, Shangrao 334000, China}

\author{Yan-Hui Zhou}
\affiliation{Quantum Information Research Center, Shangrao Normal University, Shangrao 334000, China}

\author{Chui-Ping Yang}
\thanks{yangcp@hznu.edu.cn}
\affiliation{Quantum Information Research Center, Shangrao Normal University, Shangrao 334000, China}
\affiliation{Department of Physics, Hangzhou Normal University, Hangzhou 311121, China}

\author{Franco Nori}
\thanks{fnori@riken.jp}
\affiliation{Theoretical Quantum Physics Laboratory, RIKEN, Wako-shi, Saitama 351-0198, Japan}
\affiliation{RIKEN Center for Quantum Computing (RQC), Wako-shi, Saitama 351-0198, Japan}
\affiliation{Physics Department, The University of Michigan, Ann Arbor, Michigan 48109-1040, USA}

\begin{abstract}
~~~~\\
 \begin{center}
 \textbf{\Large{Abstract}}
 \end{center}

 Non-Hermitian parity-time ($\mathcal{PT}$) and anti-parity-time ($\mathcal{APT}$)-symmetric systems exhibit novel quantum properties and have attracted increasing interest. Although many counterintuitive phenomena in $\mathcal{PT}$- and $\mathcal{APT}$-symmetric systems were previously studied, coherence flow has been rarely investigated. Here, we experimentally demonstrate single-qubit coherence flow in $\mathcal{PT}$- and $\mathcal{APT}$-symmetric systems using an optical setup. In the symmetry unbroken regime, we observe different periodic oscillations of coherence. Particularly, we observe two complete coherence backflows in one period in the $\mathcal{PT}$-symmetric system, while only one backflow in the $\mathcal{APT}$-symmetric system. Moreover, in the symmetry broken regime, we observe the phenomenon of stable value of coherence flow. We derive the analytic proofs of these phenomena and show that most experimental data agree with theoretical results within one standard deviation. This work opens avenues for future study on the dynamics of coherence in $\mathcal{PT}$- and $\mathcal{APT}$-symmetric systems.
\end{abstract}

\maketitle

\renewcommand{\figurename}{\textbf{Fig.~}}

\section{Introduction}
 Non-Hermitian Hamiltonians, satisfying parity-time ($\mathcal{PT}$) symmetry, can have real eigenvalues in the symmetry unbroken zone \cite{cm,Konotop,Ganainy}. A~$\mathcal{PT}$-symmetric~Hamiltonian~satisfies $[\hat{H},\hat{\mathcal{P}}\hat{\mathcal{T}}]=0$, with the joint parity-time operator ($\hat{\mathcal{P}}\hat{\mathcal{T}}$). $\mathcal{PT}$-symmetric non-Hermitian Hamiltonians feature \allowbreak{unconventional} properties in numerous systems ranging from classical \cite{jsa040101,fk041805,YDchong,ar167,LFeng,BPengS,HJS,LeeYC,cyju062118,bpsk394,iia053806,qjj1160} to quantum systems \cite{chgx083604,ZhangJ,Ozdemir,LiJ,TangJS,WangYT,XiaoL2,FQU,ZHBL2,iiaaf,kka}. When the Hamiltonian parameters cross the exceptional point (EP), $\mathcal{PT}$ symmetry is broken, leading to a symmetry-breaking transition. This has inspired a number of studies on many counterintuitive phenomena emerging in such systems.

  \indent Previous experiments demonstrated single-mode lasing or anti-lasing \cite{lfeng972,hhodaei975}, bistable lasing \cite{svsa18},  loss-induced transparency or lasing \cite{aguo103,BPengS}, EP-enhanced sensing \cite{ZPL,wchen192,hhodaei187}, and $\mathcal{PT}$ symmetry breaking \cite{Ozdemir,aguo103,LiJ,CER,YangW}. Moreover, recent experiments have observed: information flow in $\mathcal{PT}$-symmetric systems \cite{XiaoL2}, protection of quantum coherence in a $\mathcal{PT}$-broken superconducting circuit \cite{mnaghiloo1232}, EP-enhanced coherence and oscillation of coherence in a single ion $\mathcal{PT}$-symmetry system \cite{Chenpx}, entanglement restoration in a $\mathcal{PT}$-symmetric system using a universal circuit \cite{LongGL1}, and dynamical features of a triple-qubit system in which one qubit evolves under a local $\mathcal{PT}$-symmetric Hamiltonian \cite{LongGL2}.

 \indent Another important counterpart, anti-$\mathcal{PT}$ ($\mathcal{APT}$) symmetry, has recently attracted considerable interest. $\mathcal{APT}$ symmetry means that the system Hamiltonian is anti-commutative with the joint $\mathcal{PT}$ operator, i.e., $\{\hat{H}, \hat{\mathcal{P}}\hat{\mathcal{T}}\}=0$. $\mathcal{APT}$-symmetric systems exhibit noteworthy effects, such as balanced positive and negative index \cite{GeL}, coherent switch \cite{VVKDA}, and constant refraction \cite{YangF}. Some relevant experimental demonstrations have been realized in optics \cite{VVKDA,LiQ},~atoms \cite{jhw033811,ppengwx1139,ylchuang21969,yjy193604}, electrical circuit resonators \cite{Choi}, magnon-cavity hybrid systems \cite{jz014053}, and diffusive systems \cite{LiY}. In addition, experiments have demonstrated $\mathcal{APT}$ symmetry breaking \cite{LiY,HLZR}, simulated $\mathcal{APT}$-symmetric Lorentz dynamics \cite{LiQ}, and observed $\mathcal{APT}$ exceptional points \cite{Choi}. Moreover, information flow in an $\mathcal{APT}$-symmetric system with nuclear spins has been observed in recent experiments \cite{Wen}.

   \indent Although many counterintuitive phenomena in $\mathcal{PT}$- or $\mathcal{APT}$-symmetric systems were previously studied, the flow of coherence in $\mathcal{PT}$-symmetric systems has not been fully and thoroughly investigated. Moreover, the coherence flow in $\mathcal{APT}$-symmetric systems has not been studied either theoretically or experimentally. The study of coherence flow is interesting and meaningful because it can discover various phenomena different from Hermitian quantum mechanics and reveal the relationship between non-Hermitian systems and their environment.

   \indent In Hermitian quantum systems isolated from their environment, the coherence flow between the subsystems generally oscillates periodically over time, and the oscillation period depends on the coupling strength between the subsystems. Different from Hermitian quantum systems, most non-Hermitian physical systems typically involve gain and loss induced by the environment. In this case, the behavior of coherence flow in non-Hermitian physical systems is generally quite different from that of the coherence flow in Hermitian physical systems.

   \indent For example, the dissipative coupling between the system and the environment may disturb and even wash out quantum coherence. On the other hand, due to the gain effect, the coherence originally lost into the environment may return to the system, and thus oscillates periodically over time. By investigating the coherence flow between the system and its environment, one can obtain some important information, such as the coupling strength between systems and their environment, the type of environment (i.e., Markov or non-Markov environment), the presence of memory effects in the open quantum dynamics, etc..

  \indent In this study, we experimentally demonstrate the coherence flow of a single qubit in $\mathcal{PT}$- and $\mathcal{APT}$-symmetric systems using a simple optical setup. In the symmetry unbroken regime, we observe different periodic oscillations of coherence in both $\mathcal{PT}$- and $\mathcal{APT}$-symmetric systems. Double touch of coherence (DTC) (i.e., complete coherence backflow happening twice in one period) is revealed in the $\mathcal{PT}$-symmetric system, while only one backflow exists in the $\mathcal{APT}$-symmetric system. In addition, we observe the phenomenon of stable value (PSV) of coherence in the symmetry broken regime, which is independent of its initial state. Concretely, the coherence tends to a stable value $1/a$ in the $\mathcal{PT}$-symmetric system, but it approaches 1 in the $\mathcal{APT}$-symmetric system. We also provide the theoretical analytic proofs of these phenomena (see Supplementary Notes 1-4) and compare with previous relevant works. Our results imply that the coherence backflow and PSV are quite different for these two kinds of symmetric systems.

\begin{figure*}[!tbp]
\setlength{\belowcaptionskip}{-0.15cm}
\centerline{
\includegraphics[height=3.9cm, width=12.6cm]{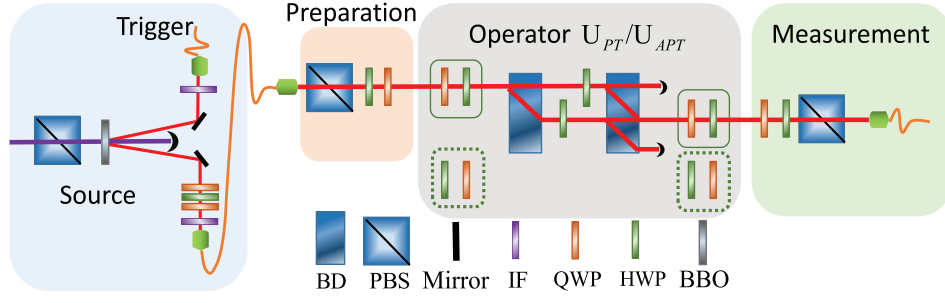}
}
\caption{\textbf{Experimental setup}. Blue area to the left: Pairs of 808 $\rm{nm}$ single photons are generated by passing a 404 $\rm{nm}$ laser light through a type-I spontaneous parametric down conversion, and using a nonlinear-barium-borate (BBO) crystal. Orange area: After photons pass through the $3$ $\rm{nm}$ interference filter (IF), one photon serves as a trigger and the other signal photon is prepared in an arbitrary linear polarization state. Grey area: Two sets of beam displacers (BDs), together with half-wave plates (HWPs) and quarter-wave plates (QWPs), are used to construct the operators $\boldsymbol{U}_{\mathcal{PT}}$ and $\boldsymbol{U}_{\mathcal{APT}}$. In the measurement part, the density matrix is constructed via quantum state tomography. PBS: polarization beam splitter.}
\label{fig1}
\end{figure*}

\section{Results}
\noindent \textbf{Principle and setup of the experiment.} A non-trivial general $\mathcal{PT}$-symmetric Hamiltonian for a single qubit takes the form \cite{LeeYC,TangJS}
\setlength\abovedisplayskip{8pt}
\setlength\belowdisplayskip{8pt}
\begin{equation}
\hat{H}_{\mathcal{PT}}=\left(\begin{array}{ll}
i\gamma & s\\
s & -i\gamma
\end{array}\right)=s(\hat{\sigma}_{x}+i a \hat{\sigma}_{z}).\label{Eq:Hpt}
\end{equation}
While a generic $\mathcal{APT}$-symmetric Hamiltonian of a single qubit can be expressed as \cite{Wen,LiY}
\begin{equation}
\hat{H}_{\mathcal{APT}}=\left(\begin{array}{ll}
\gamma & is\\
i s & -\gamma
\end{array}\right)=s( i \hat{\sigma}_{x}+a \hat{\sigma}_{z}).\label{Eq:Hapt}
\end{equation}
Here, the parameter $s>0$ is an energy scale, $a=\gamma/s>0$ is a coefficient representing the degree of non-Hermiticity, $\hat{\sigma}_{x}$ and $\hat{\sigma}_{z}$ are the standard Pauli operators. In the $\mathcal{PT}$-symmetric system, the eigenvalue of $\hat{H}_{\mathcal{PT}}$ is
\begin{equation}
 \lambda_{\mathcal{PT}}= \pm s\sqrt{1-a^{2}},
 \end{equation}
 which is an imaginary number for $a>1$ (the $\mathcal{PT}$ symmetry broken regime), while a real number for $0<a<1$ (the $\mathcal{PT}$ symmetry unbroken regime). However, in the $\mathcal{APT}$-symmetric system, the eigenvalue of $\hat{H}_{\mathcal{APT}}$ is
 \begin{equation}
  \lambda_{\mathcal{APT}}=\pm s\sqrt{a^{2}-1},
 \end{equation}
 which is an imaginary number for $0<a<1$ (the $\mathcal{APT}$ symmetry broken regime), while a real number for $a\!\!>\!\!1$ (the $\mathcal{APT}$ symmetry unbroken regime). Note that the eigenvalues of both Hamiltonians $\hat{H}_{\mathcal{PT}}$ and $\hat{H}_{\mathcal{APT}}$ are zero for $a=1$ (the exceptional point). \\
\indent For different $s$, the time evolution of quantum states under the Hamiltonian $\hat{H}_{\mathcal{PT}}$ ($\hat{H}_{\mathcal{APT}}$) follows the same rules because $s$ is an energy scale. Therefore, without loss of generality, we consider $s=1$ for both $\hat{H}_{\mathcal{PT}}$ and $\hat{H}_{\mathcal{APT}}$ \cite{WangYT,XiaoL2}. In our experiment, the non-unitary operators $\boldsymbol{U}_{\mathcal{PT}}=\mathrm{exp}(-i\hat{H}_{\mathcal{PT}}t)$ and $\boldsymbol{U}_{\mathcal{APT}}=\mathrm{exp}(-i\hat{H}_{\mathcal{APT}}t)$ are realized by \cite{XiaoL2,Stewart}
 \begin{eqnarray}
 \boldsymbol{U}_{\mathcal{PT}} =  & R_{\mathrm{HWP}}\left(\theta_{1}\right) R_{\mathrm{QWP}}\left(2\theta_{1}\right) L\left(\xi_{1}, \xi_{2}\right)\nonumber\\
  &R_{\mathrm{HWP}}\left(-\theta_{1}+\pi/4\right) R_{\mathrm{QWP}}\left(0\right),\\
 \boldsymbol{U}_{\mathcal{APT}}  =  & R_{\mathrm{QWP}}\left(0\right)R_{\mathrm{HWP}}\left(\pi/4\right) L\left(\xi_{3}, \xi_{3}\right)\nonumber \\
 &R_{\mathrm{QWP}}\left(\varphi_{1}\right)R_{\mathrm{HWP}}\left(\varphi_{2}\right),
 \end{eqnarray}
 where the loss-dependent operator
 \begin{equation}
 L\left(\xi_{i}, \xi_{j}\right)  = \left(\begin{array}{ll}
0 & \mathrm{sin} 2\xi_{i}\\
\mathrm{sin} 2\xi_{j} & 0
\end{array}\right)
 \end{equation}
 is realized by a combination of two beam displacers (BDs) and two half-wave plates (HWPs) with setting angles $\xi_{i}$ and $\xi_{j}$ (see Supplementary Note 5) \cite{XiaoL2}. Above, $R_{\textrm{HWP}}$  and $R_{\textrm{QWP}}$ are the rotation operators of HWP and QWP (quarter wave plate), respectively. Here, the setting angles $(\theta_{1},\varphi_{1},\varphi_{2},\xi_{1},\xi_{2},\xi_{3})$ depend on the initial state and are determined numerically by reversal design for each given time $t$, according to the time-evolution operators $\boldsymbol{U}_{\mathcal{PT}}$ and $\boldsymbol{U}_{\mathcal{APT}}$. \\
 \indent The dynamical evolution of the quantum states in the $\mathcal{PT}$- or $\mathcal{APT}$-symmetric system is given by \cite{kka,XiaoL2,Brody}
\begin{equation}
\boldsymbol{\rho}(t)=\frac{\boldsymbol{U}(t) \boldsymbol{\rho}(0) \boldsymbol{U}^{\dagger}(t)}{\rm Tr\left[\boldsymbol{U}(t) \boldsymbol{\rho}(0) \boldsymbol{U}^{\dagger}(t)\right]},
\end{equation}
where $\boldsymbol{U}(t)=\boldsymbol{U}_{\mathcal{PT}}(t)$ or $\boldsymbol{U}_{\mathcal{APT}}(t)$, $\boldsymbol{\rho}(0)$ is the initial density matrix, and $\boldsymbol{\rho}(t)$ is the density matrix at any given time $t$. Here, we use the $l_{1}$ norm of coherence \cite{Baumgratz,Mani} to quantify the coherence of $\boldsymbol{\rho}(t)$, i.e.,
\begin{equation}
C_{l_{1}}\left(\boldsymbol{\rho}(t)\right)=\sum_{i \neq j}\left|\boldsymbol{\rho}(t)_{i, j}\right|,\label{l1norm}
\end{equation}
where $\boldsymbol{\rho}(t)_{i,j}$ denotes the matrix element obtained from $\boldsymbol{\rho}(t)$ by deleting all diagonal elements. In the single-qubit case, Eq.~(\ref{l1norm}) is simplified as
\begin{equation}
C_{l_{1}}(\boldsymbol{\rho}(t))=\left|\boldsymbol{\rho}(t)_{1, 2}\right|+\left|\boldsymbol{\rho}(t)_{2, 1}\right|.\label{l1normsingle}
\end{equation}
Here, $\boldsymbol{\rho}(t)_{1, 2}$ and $\boldsymbol{\rho}(t)_{2, 1}$ are the two off-diagonal elements of the single-qubit density matrix.

\indent As shown in Fig.~1, our experimental setup consists of four parts (photon source, state preparation, implementation of the operator $\boldsymbol{U}_{\mathcal{PT}}$ or $\boldsymbol{U}_{\mathcal{APT}}$, and measurement). In the photon-source part, we generate heralded single photons via type-I spontaneous parametric down-conversion, with one photon serving as a trigger and the other as a signal photon (blue area). Because of the disturbance of the single-mode fiber to polarization, the signal photon needs to pass through the sandwich structure (QWP-HWP-QWP) to eliminate this influence, and then goes through various optical elements. In the orange area, we finish preparing the single-qubit arbitrary quantum state $\alpha|H\rangle+\beta e^{i\varphi}|V\rangle$ ($|\alpha|^{2}+|\beta|^{2}=1$, $\alpha, \beta \in R$) after the HWP and QWP. Before the signal photon enters the gray region, we separately prepare three initial quantum states $|H\rangle$, $\left(|H\rangle+|V\rangle\right)/\sqrt{2}$, and $\left(|H\rangle+\sqrt{3}|V\rangle\right)/2$, by appropriately choosing the rotation angles of the HWP and the QWP in the state preparation part.

 \indent The gray part has the function of simulating the $\boldsymbol{U}_{\mathcal{PT}}$ or $\boldsymbol{U}_{\mathcal{APT}}$. The loss operator $L$ can be implemented with two sets of BD and two HWPs between BDs. For the HWP along the up (bottom) path, the angle is $\xi_{1}$ $(\xi_{2})$. In order to simulate $\boldsymbol{U}_{\mathcal{PT}}$, we choose the plate combinations in the solid green wireframe. While the plates in the dotted green wireframe are used to simulate $\boldsymbol{U}_{\mathcal{APT}}$.

\indent In the measurement part (green area), the density matrix at any given time $t$ can be constructed via quantum state tomography after the signal photon passes through the gray region. Essentially, we measure the probabilities of the photon in the bases $\{|H\rangle, |V\rangle, |R\rangle=(|H\rangle-i|V\rangle)/\sqrt{2}, |D\rangle=(|H\rangle+|V\rangle)/\sqrt{2}\}$ through a combination of QWP, HWP, and PBS (polarization beam splitter), and then perform a maximum-likelihood estimation of the density matrix (tomography). The outputs are recorded in coincidence with trigger photons. The measurement of the photon source  \allowbreak{yields} a maximum of 30,000 photon counts over 3 s after the 3 nm interference filter (IF).\\


\begin{figure*}[!htbp]
\setlength{\belowcaptionskip}{0.3cm}
\centerline{
\includegraphics[height=3.9cm, width=12.6cm]{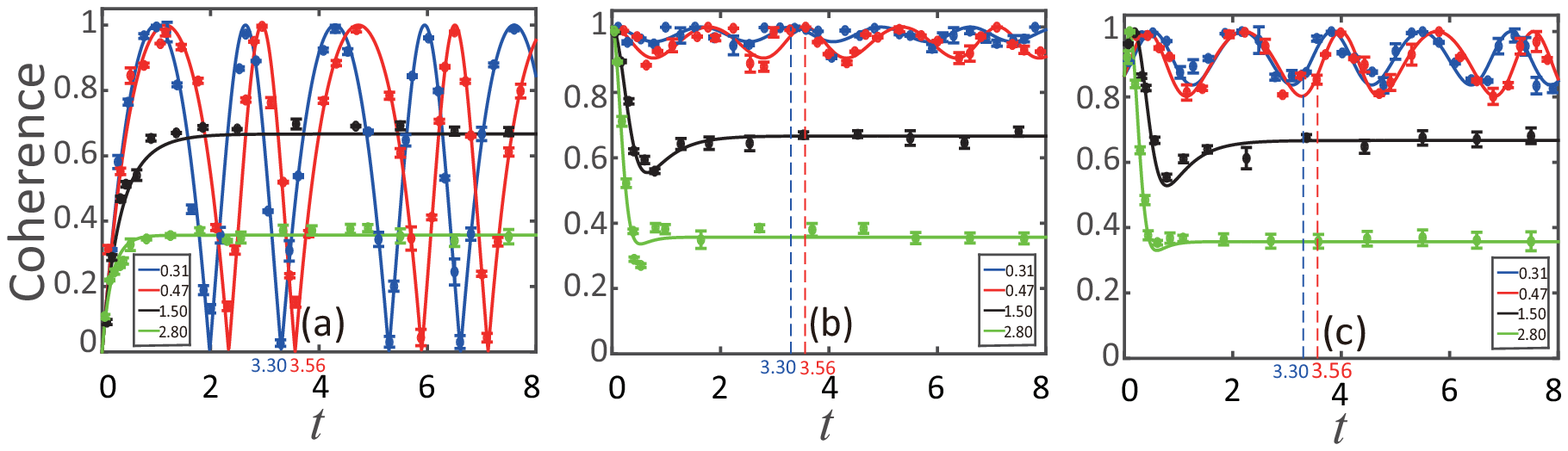}
}
\caption{\textbf{The evolution of coherence for three initial quantum states in the $\mathcal{PT}$-symmetric system.} (a) is for the initial state $|H\rangle$, (b) is for the initial state $(|H\rangle+|V\rangle)/\sqrt{2}$, (c) is for the inital state $(|H\rangle+\sqrt{3}|V\rangle)/2$. In (a, b, c), the periodic oscillation (coherence periodic backflow) happens when $a=0.31$ (blue curve) and $a=0.47$ (red curve) (the $\mathcal{PT}$ symmetry unbroken regime), while the phenomenon of stable value (PSV) of coherence occurs when $a=1.5$ (black curve) and $a=2.8$ (green curve) (the $\mathcal{PT}$ symmetry broken regime). In (a, b, c), $T=3.30$ for $a=0.31$, $T=3.56$ for $a=0.47$. In (a), $SV=0.67$ for $a=1.5$, $SV=0.36$ for $a=2.8$. In (b), $SV=0.67$ for $a=1.5$, $SV=0.36$ for $a=2.8$. In (c), $SV=0.67$ for $a=1.5$, $SV=0.36$ for $a=2.8$. ``$SV$'' means stable value. All curves are theoretical results while the dots are the experimental data. The  experimental errors of one standard deviation (1 SD) are estimated from the statistical variation of photon counts, which satisfy the Poisson distribution.}
\label{fig2}
\end{figure*}
 \noindent \textbf{{Experimental results.}} Figure 2(a, b, c) demonstrate the time-evolution dynamics of the coherence of three initial quantum states $|H\rangle$, $(|H\rangle+|V\rangle)/\sqrt{2}$, and $(|H\rangle+\sqrt{3}|V\rangle)/2$ in the $\mathcal{PT}$-symmetric system. Coherence varies over time $t$ for: (i) $a=0.31$ (blue curve), $a=0.47$ (red curve) $(0<a<1)$; and (ii) $a=1.5$ (blue curve), $a=2.8$ (green curve) $(a>1)$. For $0<a<1$ (the $\mathcal{PT}$ symmetry unbroken regime), coherence oscillates (see blue and red curves) suggesting a coherence complete recovery and backflow. There are two complete  backflows of coherence in one  period, i.e., double touch of coherence (DTC), which is observed in our experiment and agrees with our theoretical results (see Supplementary Note 3). However, for $a>1$ (the $\mathcal{PT}$ symmetry broken regime), a PSV of coherence occurs (see dark and green curves). Extracted from the experimental data, the recurrence time fits the theoretical value given by
 \begin{equation}
 T_{\mathcal{PT}}=\frac{\pi}{\sqrt{1-a^{2}}},
 \end{equation}
 and the stable value for the PSV agrees well with the theoretical value $1/a$ (see Supplementary Note 1).

\indent For the same three initial quantum states in the $H_{\mathcal{APT}}$ case, the dynamical characteristics of coherence are shown in Fig.~3, where Figs.~3(a, b, c) are respectively for the initial states $|H\rangle$, $(|H\rangle+|V\rangle)/\sqrt{2}$ and $(|H\rangle+\sqrt{3}|V\rangle)/2$. In contrast to the $H_{\mathcal{PT}}$ case, coherence oscillations occur for $a>1$ (the $\mathcal{APT}$ symmetry unbroken regime), while PSV occurs for $0<a<1$ (the $\mathcal{APT}$ symmetry broken regime), as verified in Figs.~3(a, b, c). Different from the PSV in the $\mathcal{PT}$-symmetric system, the stable value for the PSV in the $\mathcal{APT}$-symmetric system is 1 (see blue and red curves). Figure~3(b) shows that the saturated coherence does not change over time $t$ for any value of $a$. As demonstrated in Figs.~3(a, c), there exits only a single backflow in one period (see dark and green curves), i.e. the DTC phenomenon does not occur in the $\mathcal{APT}$-symmetric system (the theoretical proof is in Supplementary Note 4).
\begin{figure*}[!htbp]
\setlength{\belowcaptionskip}{0cm}
\centerline{
\includegraphics[height=3.9cm, width=12.6cm]{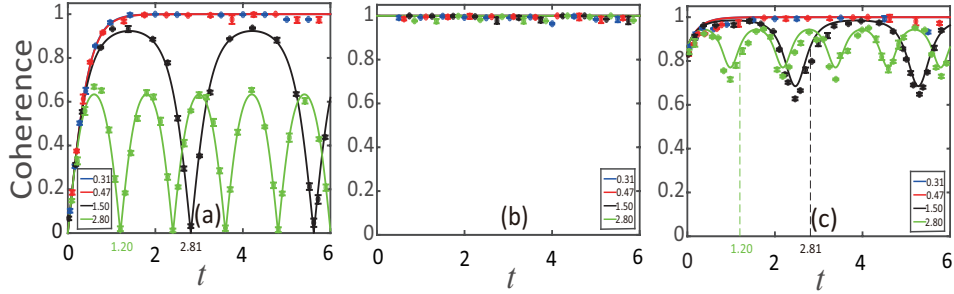}
}
\caption{\textbf{The evolution of coherence for three initial quantum states in the $\mathcal{APT}$-symmetric system}. (a) is for the initial state $|H\rangle$, (b) is for the initial state $(|H\rangle+|V\rangle)/\sqrt{2}$, while (c) is for the initial state $(|H\rangle+\sqrt{3}|V\rangle)/2$. In (a) and (c), the coherence evolution exhibits periodic backflow when $a=1.5$ (black curve) and $a=2.8$ (green curve) (the $\mathcal{APT}$ symmetry unbroken regime), while the PSV of coherence occurs when $a=0.31$ (blue curve) and $a=0.47$ (red curve) (the $\mathcal{APT}$ symmetry broken regime). In (b), coherence is conserved and does not change over time $t$, independent of $a$. In (a) and (c), $T=2.81$ for $a=1.5$, $T=1.20$ for $a=2.8$.
All curves are theoretical results while the dots are the experimental data. The  experimental errors of 1 SD are estimated from the statistical variation of photon counts, which satisfy the Poisson distribution.}
\label{fig3}
\end{figure*}

 \indent The oscillating period observed in the experiment is consistent with the theoretical value given by
 \begin{equation}
 T_{\mathcal{APT}}=\frac{\pi}{\sqrt{a^{2}-1}},
  \end{equation}
  and the stable value for the PSV observed in the experiment is in a good agreement with the theoretical value 1 (see Supplementary Note 2).

  \begin{figure}[!htbp]
\setlength{\belowcaptionskip}{0cm}
\centerline{
\includegraphics[width=0.8\linewidth]{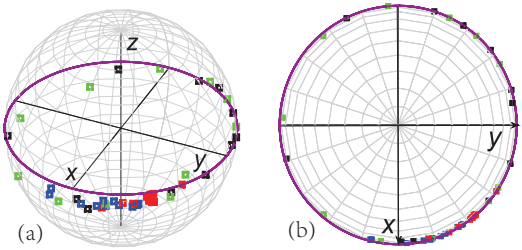}
}
\caption{\textbf{The trajectory evolution of the initial quantum state $\left(|H\rangle+|V\rangle\right)/\sqrt{2}$ shown by a circle on the Bloch sphere in the $\mathcal{APT}$-symmetric system for different values of $a$}. (a) 3D and (b) overhead views. All curves are theoretical results, while squares represent the measured quantum states in our experiment for different evolution times. Different-color squares represent different values of $a$ (blue square: $a=0.31$, red square: $a=0.47$, black square: $a=1.5$, green square: $a=2.8$). Note that the coherences of quantum states on the purple circle are the same. No error bar is plotted because it is difficult to show the error of a quantum state on the Bloch sphere.}
\label{fig4}
\end{figure}

\indent To better understand Fig.~3(b), we plot Fig.~4, which shows the trajectory evolution of the initial quantum state $\left(|H\rangle+|V\rangle\right)/\sqrt{2}$ on the Bloch sphere in the $\mathcal{APT}$-symmetric system. Figure~4 shows that the evolved quantum state travels over time along the outer edge of the $XY$ plane, which is independent of $a$. Thus, it can intuitively reflect why the coherence of the quantum state [shown in Fig.~3(b)] remains unchanged during the time evolution.

\section{Discussion}
 \indent Our setup provides  a simple platform to  investigate both  $\mathcal{PT}$- and  $\mathcal{APT}$-symmetric systems. First, the gain and loss, associated with dissipative coupling between the system and environment, can be readily simulated with optical elements. By selecting the appropriate combination of optical elements with adjustable angles, both $\mathcal{PT}$- and $\mathcal{APT}$-symmetric systems can be realized with this setup. Second, our setup can be used to demonstrate the dynamics of $\mathcal{PT}$- and $\mathcal{APT}$-symmetric systems for each given  evolution time $t$, by performing the corresponding nonunitary gate operations on the initial states. The dynamics of $\mathcal{PT}$- and $\mathcal{APT}$-symmetric systems for each given evolution time $t$ is stable and the coherence time of photons is long enough, thus one can accurately extract the critical information from the nonunitary dynamics.

 \indent Let us briefly recall the difference between Rabi oscillations and coherence flow oscillations. In our work, we only consider a single qubit, with the usual two logical states $|0\rangle$ and $|1\rangle$. Rabi oscillations refer to the dynamical evolution of the population probability of the logical state $|0\rangle$ or $|1\rangle$ of the qubit. For example, this occurs when the qubit is placed inside a cavity and the cavity-qubit coupling is sufficiently strong, so there is an exchange of energy between the qubit and the photons bouncing back and forth many times inside the cavity. Also, Rabi oscillations occur when a classical driving field is applied to a qubit, where there is an exchange of energy between the qubit and the drive, and the Rabi frequency is proportional to the applied driving field amplitude. On the other hand, for a single qubit, the coherence of quantum states is defined as the sum of the two off-diagonal elements of the single-qubit density matrix, according to Eq.~(\ref{l1norm}). Coherence flow oscillations refer to the oscillations of the coherence of quantum states. Different from Rabi oscillations, coherence flow oscillations do not require the qubit to exchange energy with photons located in a cavity or exchange energy with a classical pulse. Clearly, Rabi oscillations and coherence flow oscillations are completely different notions.

 \indent Now let us make a brief comparison with previous works \cite{XiaoL2,kka,Chenpx,Wen}, which are most relevant to this work:

\indent (i) A theoretical and experimental research on the dynamics of coherence under $\mathcal{PT}$-symmetric system has been recently presented by Wang \emph{et al.} \cite{Chenpx} in a single-ion system. There, the coherence evolution was discussed by the time average of the coherence and the diagonal element (e.g., $\rho_{00}$) of the quantum state density matrix \cite{Chenpx}. In our work, we provide a simple platform to demonstrate  $\mathcal{PT}$- and $\mathcal{APT}$-symmetric systems in experiments. We discuss the  $l_1$  norm of the coherence (i.e., the summation of off-diagonal elements of the density matrix), which is quite different from the time average of the coherence. Furthermore, we theoretically predict some phenomena (the DTC and PSV phenomena) which were not reported by Wang \emph{et al.} \cite{Chenpx}, and give an experimental demonstration in a linear optical system.

\indent (ii) In previous works on information flow \cite{kka,XiaoL2}, the trace distance
\begin{equation}
D\left(\rho_1(t),~\rho_2(t)\right)=\frac{1}{2}\rm Tr|\rho_1(t)-\rho_2(t)|
 \end{equation}
 was introduced to characterize the information flow; while in our work the $l_1$ norm of the coherence, described by Eq.~(\ref{l1norm}), is introduced to characterize the coherence flow. The concept of coherence flow is different from information flow. Second, the trace distance $D\left(\rho_1(t),~\rho_2(t)\right)$ generally measures the distinguishability of  two quantum states, while the $l_1$ norm of coherence quantifies the coherence of a quantum state. In this sense, the physical meaning of the coherence flow is different from that of the information flow. Third, the physical phenomena revealed by the coherence flow and the information flow are not exactly the same.   For   example,   we   found   that  in the $\mathcal{PT}$-symmetry unbroken regime, there are two complete coherence backflows in one period; while, the previous works  \cite{kka,XiaoL2} showed that in the $\mathcal{PT}$-symmetry unbroken regime, there exists only one information flow within one period.  Last, quantum coherence is an intriguing property of quantum states, which is a key resource in quantum computing, quantum communication, and quantum metrology; and the $\mathcal{PT}$/$\mathcal{APT}$ systems have attracted considerable interest. Thus, we believe that it is significant to study the evolution of coherence of quantum states in $\mathcal{PT}$ and $\mathcal{APT}$ systems.

\indent (iii) It is obvious that our work differs from the work by Wen \emph{et al.} \cite{Wen}. Their work studied the information flow in an $\mathcal{APT}$-symmetry system, which is different from the coherence flow; while, the present work focuses on the coherence flow.

\section{Conclusions}
\indent In summary, we have experimentally demonstrated the coherence flow in both $\mathcal{PT}$- and $\mathcal{APT}$-symmetric systems by using a single-photon qubit. In this paper, the DTC phenomenon in one period in the $\mathcal{PT}$-symmetric unbroken regime has been demonstrated, which however does not occur in the $\mathcal{APT}$-symmetric system. Moreover, the PSV has been observed in the $\mathcal{PT}$/$\mathcal{APT}$-symmetric broken regime, which is independent of the initial state. As an extension of this work, we have numerically simulated the dynamics of coherence for two-qubit $\mathcal{PT}$/$\mathcal{APT}$ systems (for details, see Supplementary Note 6). The simulations show that for both two-qubit $\mathcal{PT}$/$\mathcal{APT}$ systems, there exist different periodic oscillations of coherence (including one coherence backflow, two coherence backflows, and multiple coherence backflows in one period) in the unbroken regime; while there exists PSV in the broken regime, which is independent of the initial state. Our work merits future study on the multi-qubit coherence flow in $\mathcal{PT}$-and $\mathcal{APT}$-symmetric systems, which is left as an open question.

\section{Methods}
\noindent \textbf{Device parameters.} The photon-source system of the single-qubit, the pump laser power is 130 $\rm{mW}$.  In the state preparation part, three initial quantum states $|H\rangle$, $\left(|H\rangle+|V\rangle\right)/\sqrt{2}$, and $\left(|H\rangle+\sqrt{3}|V\rangle\right)/2$ corresponding angles of HWP are $0^{\circ}$, $22.5^{\circ}$ and $30^{\circ}$, respectively.~In~the~measure part, the four bases $\left\{|H\rangle, |V\rangle, |R\rangle=\left(|H\rangle-i|V\rangle\right)/\sqrt{2}, |D\rangle=\left(|H\rangle+|V\rangle\right)/\sqrt{2}\right\}$ corresponding angles of QWP-HWP are $\left(0^{\circ},0^{\circ}\right)$, $\left(0^{\circ},45^{\circ}\right)$, $\left(0^{\circ},22.5^{\circ}\right)$ and $\left(45^{\circ},22.5^{\circ}\right)$, respectively.

\noindent \textbf{Analysis of experimental imperfections.}~Due~to~the accuracy of the rotation angle, and the imperfection of the interference visibility between BDs, several points of experiment data do not fit well with our theoritical data. To solve this problem, we improve the extinction ratio of interference between BDs for a high interference visibility. Instead of manual adjustment, we use motorized precision rotation mount to ensure the higher accuracy of the plate rotation angle. Meanwhile, the experimental errors are estimated from  the statistical variation of photon counts, which satisfy the Poisson distribution.

\section*{Data availability}
The data that support the findings of this study are available from the corresponding authors upon reasonable request.

\section*{Code availability}
The code used for simulations is available from the corresponding authors upon reasonable request.

\section*{Acknowledgments}
 This work was partly supported by the Jiangxi Natural Science Foundation (20192ACBL20051),
 the National Natural Science Foundation of China (NSFC) (11074062, 11374083, 11774076, 11804228), and the Key-Area Research and Development Program of GuangDong province (2018B030326001). F.N. is supported in part by: Nippon Telegraph and Telephone Corporation (NTT) Research, the Japan Science and Technology Agency (JST) [via the Quantum Leap Flagship Program (Q-LEAP), the Moonshot R\&D Grant Number JPMJMS2061, and the Centers of Research Excellence in Science and Technology (CREST) Grant No. JPMJCR1676], the Japan Society for the Promotion of Science (JSPS) [via the Grants-in-Aid for Scientific Research (KAKENHI) Grant No. JP20H00134 and the JSPS-RFBR Grant No. JPJSBP120194828], the Army Research Office (ARO) (Grant No.~W911NF-18-1-0358), the Asian Office of Aerospace Research and Development (AOARD) (via Grant No. FA2386-20-1-4069), and the Foundational Questions Institute Fund (FQXi) via Grant No. FQXi-IAF19-06.

\section*{Author contributions}
  Y.-L. F. performed the experiment and analyzed the data with the assistance of Y.Z., J.-L.Z and C.-P. Y., Y.Z. proposed the experiment and designed experimental scheme. J.-L. Z. provided the theoretical analytic proofs of our results. C.-P. Y. and F. N. directed the project. J.-L. Z, Y. Z. D.-X. C Y.-H. Z., Q.-C. W. C.-P Y. and F. N. provided theoretical support. Y.Z., J.-L.Z., Q.-C.W., C.-P.Y., and N.F. wrote the manuscript with feedback from all authors.

  \section*{Competing interests}
   The authors declare no competing interests

\clearpage
\begin{widetext}
\section*{\large{Supplementary Information for:\setlength{\baselineskip}{22pt}\\
Experimental demonstration of coherence flow in $\mathcal{PT}$- and anti-$\mathcal{PT}$-symmetric systems}}
\addcontentsline{toc}{section}{Appendices}\markboth{APPENDICES}{}
\renewcommand{\figurename}{{Supplementary Figure~}}
\setcounter{equation}{0}
\setcounter{subsection}{0}
\setcounter{figure}{0}
\renewcommand{\theequation}{S\arabic{equation}}
\renewcommand{\thefigure}{\arabic{figure}}

 \begin{center}
 \makeatletter
 \renewcommand{\section}{\textbf{Supplementary Note 1:~}}
\section{\textbf{ Phenomenon of stable value and the period of coherent evolution in $\mathcal{PT}$-symmetric systems}}
 \end{center}

 Let us first consider the $\mathcal{PT}$-symmetric non-Hermitian Hamiltonian shown in Eq.~(1) of the main text. The evolution of quantum states in $\mathcal{PT}$-symmetric systems is described by the time-evolution operator $U_{\mathcal{PT}}=\exp(-i\hat{H}_{\mathcal{PT}}t)$:
 \begin{eqnarray}
 \boldsymbol{U}_{\mathcal{PT}}(t)&=&\exp(-\emph{i}\hat{H}_{\mathcal{PT}}t)\nonumber\\
        &=&\exp\left[-\emph{i}s(\sigma_{x}+\emph{i} a \sigma_{z})t\right]\nonumber \\
        &=&\exp\left[ s\left(\begin{array}{cc}
                    a & -\emph{i}\\
                    -\emph{i}   & -a
                  \end{array}
             \right)t\right]\nonumber\\
          & =&\left(\begin{array}{cc}
                    A-B & -\emph{i}C\\
                    -\emph{i}C   & A+B
                  \end{array}
             \right).\label{Eq:Upt}
 \end{eqnarray}
 Here $A$, $B$ and $C$ are given by:

 (i) for $0<a<1$,
 \begin{equation}
  A=\cos \left(\omega_1 st\right),~~B=-\frac{a}{\omega_1}\sin\left(\omega_1 st\right),~~C=\frac{1}{\omega_1}\sin\left(\omega_1 st\right),\label{Eq:PTABC}
 \end{equation}
where $\omega_1=\sqrt{1-a^2}>0$.

 (ii) for $a>1$,
 \begin{equation}
 A=\cosh\left(\omega_2 st\right),~~B=-\frac{a}{\omega_2}\sinh\left(\omega_2 st\right),~~C=\frac{1}{\omega_2 }\sinh\left(\omega_2 st\right),\label{Eq:PTABC2}
 \end{equation}
where $\omega_2=\sqrt{a^2-1}>0$.

 In general, the initial state is $|\phi\rangle=\alpha|H\rangle+\beta e^{i\varphi}|V\rangle$, where $\alpha,\beta \in \left[0,1\right]$, $\alpha^2+\beta^2=1$ and $\varphi \in \left[0,2\pi\right]$. The time-evolved state is expressed as:
\begin{eqnarray}
  |\phi(t)\rangle & = & \frac{\boldsymbol{U}_{\mathcal{PT}}|\phi\rangle}{\|\boldsymbol{U}_{\mathcal{PT}}|\phi\rangle\|}\nonumber\\
 & = &\frac{1}{\sqrt{M}}\left(
                \begin{array}{c}
                  \alpha(A-B)-iC\beta e^{i\varphi}\\
                  (A+B)\beta e^{i\varphi}-iC\alpha
                \end{array}
              \right),\label{Eq:Phipt1}
\end{eqnarray}
where $\| |\cdot\rangle\|=\sqrt{\Tr\left(|\cdot\rangle \langle \cdot |\right)}$ denotes the normalization coefficient, and
\begin{equation}
 M=\| |\cdot\rangle\|^2=A^2+B^2+C^2+2AB\left(\beta^2-\alpha^2\right)-4\alpha\beta BC\sin \varphi.
 \end{equation}
 Thus, the coherence of the state $|\phi(t)\rangle$ is given by:
 \begin{equation}
 C_{l_1} \left(|\phi(t)\rangle\right) =\frac{2\left\{\left[\alpha^2(A-B)^2+C^2\beta^2+2\alpha\beta (AC-BC)\sin\varphi\right]\left[C^2\alpha^2+(A+B)^2\beta^2-2\alpha\beta (AC+BC)\sin\varphi\right]\right\}^{1/2}}{A^2+B^2+C^2+2AB\left(\beta^2-\alpha^2\right)-4\alpha\beta BC\sin \varphi}.\label{Eq:CPTphi}
\end{equation}

Let us first consider the case when $0<a<1$ (i.e., the $\mathcal{PT}$-symmetric-unbroken regime). In this case, $A$, $B$, and $C$ are given by Eq.~(\ref{Eq:PTABC}). After inserting Eq.~(\ref{Eq:PTABC}) into  Eq.~(\ref{Eq:CPTphi}), a simple calculation gives:
\begin{equation}
 C_{l_1} \left(|\phi(t)\rangle\right)=\frac{2\sqrt{m_1}}{1+m_1},\label{Eq:CPTphi2}
\end{equation}
where $m_1=x_1/y_1$, with $x_1$ and $y_1$ given below:
\begin{eqnarray}
 x_1=\frac{1}{\omega_1^2}\left[\alpha^2\left(\omega_1^2\cos 2\theta_1 +a\omega_1\sin 2\theta_1 \right)+\frac{1-\cos 2\theta_1}{2}+\alpha\beta\sin\varphi\left(\omega_1\sin 2\theta_1+a(1-\cos 2\theta_1)\right)\right],\nonumber\\
 y_1=\frac{1}{\omega_1^2}\left[\beta^2\left(\omega_1^2\cos 2\theta_1-a\omega_1\sin 2\theta_1\right)+\frac{1-\cos 2\theta_1}{2}-\alpha\beta\sin\varphi\left(\omega_1\sin 2\theta_1-a(1-\cos 2\theta_1)\right)\right].\label{Eq:PTCxy1}
\end{eqnarray}
Here $\theta_1=\omega_1 st$. From Eq.~(\ref{Eq:CPTphi2}) and Eq.~(\ref{Eq:PTCxy1}), one can see that $ C_{l_1} \left(|\phi(t)\rangle\right)$ is a function of $\sin 2\theta_1$ and $\cos 2\theta_1$. Thus, the period of $C_{l_1}\left(|\phi\rangle\right)$ is the same as that of $\sin 2\theta_1$ or $\cos 2\theta_1$. Note that $\sin 2\theta_1$ ($\cos 2\theta_1$) can be written as $\sin 2\omega_1 st$ ($\cos 2\omega_1 st$) with $\omega_1=\sqrt{1-a^2}$. Therefore, the period of coherent evolution in $\mathcal{PT}$-symmetric systems is:
 \begin{equation}
  T_{\mathcal{PT}}=\frac{2\pi}{2\omega_1 s}=\frac{\pi}{s\sqrt{1-a^2}}.\label{Eq:Tpt}
 \end{equation}

Now, let us consider the case $a>1$ (i.e., the $\mathcal{PT}$-symmetric-broken regime). In this situation, $A$, $B$, and $C$ are given by Eq.~(\ref{Eq:PTABC2}). Substitution of Eq.~(\ref{Eq:PTABC2}) into Eq.~(\ref{Eq:CPTphi}) gives:

 \begin{equation}
 C_{l_1}\left(|\phi(t)\rangle\right) =\frac{2\sqrt{m_2}}{1+m_2},\label{Eq:CPTphi3}
\end{equation}
where $m_2=x_2/y_2$, with $x_2$ and $y_2$ given by
\begin{eqnarray}
 x_2=\alpha^2\left(\cosh\theta_2+\frac{a}{\omega_2}\sinh\theta_2\right)^2+\frac{1}{\omega_2^2}\beta^2\sinh^2\theta_2+ 2\alpha\beta\left(\frac{1}{\omega_2}\cosh\theta_2\sinh\theta_2+\frac{a}{\omega_2}\sinh^2\theta_2\right)\sin\varphi,\nonumber\\
  y_2=\beta^2\left(\cosh\theta_2-\frac{a}{\omega_2}\sinh\theta_2 \right)^2+\frac{1}{\omega_2^2}\alpha^2\sinh^2\theta_2- 2\alpha\beta\left(\frac{1}{\omega_2}\cosh\theta_2\sinh\theta_2-\frac{a}{\omega_2^2}\sinh^2\theta_2\right)\sin\varphi.\label{Eq:PTCxy2}
\end{eqnarray}
Here $\theta_2=\omega_2 st$. When $t \rightarrow \infty$, $\cosh \theta_2 \thicksim \sinh\theta_2\rightarrow \infty$. Thus, it is straightfordward to find from Eqs.~(\ref{Eq:CPTphi3}) and (\ref{Eq:PTCxy2}) that:
\begin{eqnarray}
 \lim_{t\rightarrow \infty}  C_{l_1} \left(|\phi(t)\rangle\right)  &=&\frac{2\left\{a^2+\omega_2^2(\alpha^2-\beta^2)^2+2a\omega_2(\alpha^2-\beta^2) +4\alpha\beta\sin\varphi\left[a+\omega_2(\alpha^2-\beta^2)\right]+4\alpha^2\beta^2\sin^2\varphi\right\}^{1/2}}{2a^2+2a\omega_2\left(\alpha^2-\beta^2\right)+4a\alpha\beta\sin\varphi}\nonumber\\
 &=&\frac{2\left[a+\omega_2 (\alpha^2-\beta^2)+2\alpha\beta\sin\varphi\right]}{2a^2+2a\omega_2\left(\alpha^2-\beta^2\right)+4a\alpha\beta\sin\varphi}\nonumber\\
 &=&\frac{1}{a}.\label{Eq:PSVPT}
\end{eqnarray}
Equation~(\ref{Eq:PSVPT}) shows that the phenomenon of stable value (PSV) of coherence occurs after a long time evolution; that is, the coherence tends to a stable value $1/a$, which is independent of the initial states.\\

 \begin{center}
 \makeatletter
 \renewcommand{\section}{\textbf{Supplementary Note 2:~}}
\section{\textbf{ Phenomenon of stable value and the period of coherent evolution in anti-$\mathcal{PT}$-symmetric systems}}
 \end{center}

 Let us now consider the anti-$\mathcal{PT}$($\mathcal{APT}$)-symmetric non-Hermitian Hamiltonian in Eq.~(2)  of the main text. The evolution of the quantum states in $\mathcal{APT}$-symmetric systems is governed by the operator $U_{\mathcal{APT}}=\exp(-i\hat{H}_{\mathcal{APT}}t)$:
 \begin{eqnarray}
   \boldsymbol{U}_{\mathcal{APT}}(t) & = & \exp(-i\hat{H}_{\mathcal{APT}}t)\nonumber\\
                           & = & \exp\left[-is(i \sigma_{x}+a \sigma_{z})t\right]\nonumber\\
     & = &\exp\left[ s\left(\begin{array}{cc}
                    -ia & 1\\
                    1    & ia
                  \end{array}
             \right)t\right]\nonumber\\
     & = &\left(\begin{array}{cc}
                    A+iB & C\\
                    C    & A-iB
                  \end{array}
             \right).\label{Uapt}
 \end{eqnarray}
Here $A$, $B$ and $C$ are given by:

(i) for $a>1$,
\begin{equation}
 A=\cos\left(\omega_3 st\right),~~B=-\frac{a}{\omega_3}\sin\left(\omega_3 st\right)t,~~C=\frac{1}{\omega_3}\sin\left(\omega_3 st\right),\label{Eq:APTABC}
\end{equation}
where $\omega_3=\sqrt{a^2-1} >0$.

(ii) for $0<a<1$,
\begin{equation}
A=\cosh\left(\omega_4 st\right),~~B=-\frac{a}{\omega_4}\sinh\left(\omega_4 st\right),~~C=\frac{1}{\omega_4}\sinh\left(\omega_4 st\right),\label{Eq:APTABC2}
\end{equation}
where $\omega_4=\sqrt{1-a^2} >0$.

 In general, the initial state is $|\phi\rangle=\alpha|H\rangle+\beta e^{i\varphi}|V\rangle$. The time-evolved state is given by:
\begin{eqnarray}
  |\phi(t)\rangle & = & \frac{\boldsymbol{U}_{\mathcal{APT}}|\phi\rangle}{\|\boldsymbol{U}_{\mathcal{APT}}|\phi\rangle\rangle\|}\nonumber\\
 & = &\frac{1}{\sqrt{M}}\left(
                \begin{array}{c}
                  \alpha(A+Bi) + C\beta e^{i\varphi}\\
                  (A-Bi)\beta e^{i\varphi}+C\alpha
                \end{array}
              \right),\label{Eq:Phiapt1}
\end{eqnarray}
where $M=A^2+B^2+C^2+ 4C\left(A\cos\varphi+B\sin\varphi\right) \alpha\beta$. The coherence of $|\phi(t)\rangle$ is:
 \begin{equation}
 C_{l_1}\left(|\phi(t)\rangle\right) =\frac{2\left\{\left[(A\alpha+C\beta\cos\varphi)^2+(B\alpha+C\beta\sin\varphi)^2\right]\left[\left((A\cos\varphi+B\sin\varphi)\beta+ C\alpha\right)^2+(A\sin\varphi-B\cos\varphi)^2\beta^2\right]\right\}^{1/2}}{A^2+B^2+C^2+ 4C\left(A\cos\varphi+B\sin\varphi\right) \alpha\beta}.\label{Eq:CAPT1}
\end{equation}

 Let us first consider the case $a>1$ (i.e., the $\mathcal{APT}$-symmetric-unbroken regime). In this case, $A$, $B$ and $C$ are given by Eq.~(\ref{Eq:APTABC}).
After inserting Eq.~(\ref{Eq:APTABC}) into Eq.~(\ref{Eq:CAPT1}), we obtain:
 \begin{equation}
 C_{l_1}\left(|\phi(t)\rangle\right) =\frac{2\sqrt{m_3}}{1+m_3},\label{Eq:CAPT2}
 \end{equation}
where $m_3=x_3/y_3$, with $x_3$ and $y_3$ given below:
\begin{eqnarray}
x_3=\frac{1}{\omega_3^2}\left[\omega_3^2\alpha^2+\alpha\beta\omega_3\cos\varphi\sin 2\theta_3+\frac{1-\cos 2\theta_3}{2}\left(1-a\alpha\beta\sin\varphi\right)\right],\nonumber\\
y_3=\frac{1}{\omega_3^2}\left[\omega_3^2\beta^2+\alpha\beta\omega_3\cos\varphi\sin 2\theta_3+\frac{1-\cos 2\theta_3}{2}\left(1-a\alpha\beta\sin\varphi\right)\right].\label{Eq:APTxy1}
\end{eqnarray}
Here $\theta_3=\omega_3 st$. Based on Eq.~(\ref{Eq:CAPT2}) and Eq.~(\ref{Eq:APTxy1}), one sees that $C_{l_1}\left(|\phi(t)\rangle\right)$ is a function of $\sin 2\theta_3$ and $\cos 2\theta_3$; that is, $\sin 2\omega_3 st$ and $\cos 2\omega_3 st$. Hence, the period of coherent evolution in $\mathcal{APT}$-symmetric systems is:
 \begin{equation}
  T_{\mathcal{APT}}=\frac{2\pi}{2\omega_3 s}=\frac{\pi}{s\sqrt{a^2-1}}.\label{Eq:Tapt}
 \end{equation}

Let us now consider the case of $0<a<1$ (i.e., the $\mathcal{APT}$-symmetric-broken regime). In this situation, $A$, $B$ and $C$ are given by Eq.~(\ref{Eq:APTABC2}). Substitution of Eq.~(\ref{Eq:APTABC2}) into Eq.~(\ref{Eq:CAPT1}) leads to:
 \begin{eqnarray}
  C_{l_1}\left(|\phi(t)\rangle\right)=\frac{2\sqrt{m_4}}{1+m_4},\label{Eq:APTH2}
 \end{eqnarray}
 where  $m_4=x_4/y_4$, with $x_4$ and $y_4$ given below:
 \begin{eqnarray}
 \lefteqn{x_4=\frac{1}{\omega_4^2}\left[\left(\omega_4^2\cosh^2\theta_4+a^2\sinh^2\theta_4\right)\alpha^2+\beta^2\sinh^2\theta_4 +2\alpha\beta\sinh\theta_4\left(\omega_4\cos\varphi\cosh\theta_4-a\sinh\theta_4\sin\varphi\right)\right],}\hspace*{470pt}\nonumber\\
  \lefteqn{y_4=\frac{1}{\omega_4^2}\left[\left(\omega_4^2\cosh^2\theta_4+a^2\sinh^2\theta_4\right)\beta^2+\alpha^2\sinh^2\theta_4 +2\alpha\beta\sinh\theta_4\left(\omega_4\cos\varphi\cosh\theta_4-a\sinh\theta_4\sin\varphi\right)\right].}\hspace*{470pt}\label{Eq:APTxy2}
 \end{eqnarray}
  Here $\theta_4=\omega_4 st$. When $t \rightarrow \infty$, $\cosh \theta_4 \thicksim \sinh\theta_4\rightarrow \infty$. Thus, it follows from Eq.~(\ref{Eq:APTxy2}) that:
 \begin{equation}
  x_4 \thicksim y_4 \thicksim \frac{1+2\alpha\beta(\sqrt{1-a^2}\cos\varphi-a\sin\varphi)}{1-a^2}\sinh^2\theta_4.
 \end{equation}
 Accordingly, it follows from Eq.~(\ref{Eq:APTH2}) that:
 \begin{eqnarray}
  \lim_{t\rightarrow \infty}C_{l_1}\left(|\phi(t)\rangle\right)   &=& \lim_{t \rightarrow \infty}\frac{2\sqrt{x_4/y_4}}{1+x_4/y_4}\nonumber\\
   &=&1. \label{Eq:CAPTphiinf}
 \end{eqnarray}
 Equation~(\ref{Eq:CAPTphiinf}) shows that the phenomenon of stable value (PSV) of coherence occurs after a long time evolution; that is, the coherence tends to $1$, which is independent of the initial states.\\

\clearpage
 \begin{center}
 \makeatletter
 \renewcommand{\section}{\textbf{Supplementary Note 3:~}}
\section{\textbf{ Proof for the characteristics of each backflow in the $\mathcal{PT}$-symmetric-unbroken regime}}
 \end{center}

  For an arbitrary initial state $|\phi\rangle=\alpha |H\rangle + \beta e^{i\varphi}|V\rangle$, the coherence of the evolved state $|\phi(t)\rangle$ in the $\mathcal{PT}$-symmetric unbroken regime is given by Eq.~(\ref{Eq:CPTphi2}). According to Eq.~(\ref{Eq:CPTphi2}), the derivative of $ C_{l_1} \left(|\phi(t)\rangle\right)$ can be decomposed into
\begin{equation}
 \frac{dC_{l_1} \left(|\phi(t)\rangle\right)}{dt}=\frac{dC_{l_1} \left(|\phi(t)\rangle\right)}{dm_1}\times \frac{dm_1}{d\theta_1}\times \frac{d\theta_1}{dt}.
\end{equation}
Because of $\frac{d\theta_1}{dt}=\omega_1 s>0$, the condition for $\frac{dC_{l_1} \left(|\phi(t)\rangle\right)}{dt}=0$ turns into:
\begin{equation}
\frac{dC_{l_1} \left(|\phi(t)\rangle\right)}{dm_1}=0\label{Eq:PTdCdm}
\end{equation}
or
\begin{equation}
\frac{dm_1}{d\theta_1}=0.\label{Eq:PTdmdthe0}
\end{equation}

First, we consider the case of $\frac{dC_{l_1} \left(|\phi(t)\rangle\right)}{dm}=0$. According to Eq.~(\ref{Eq:CPTphi2}), we have
\begin{equation}
 \frac{dC_{l_1} \left(|\phi(t)\rangle\right)}{dm_1}=\frac{1-m_1}{(1+m_1)^2\sqrt{m_1}}=0.\label{Eq:dcdm}
\end{equation}
Because of $m_1=x_1/y_1$, it follows from Eq.~(\ref{Eq:PTCxy1}) that:
\begin{equation}
 m_1=\frac{\alpha^2\left(\omega_1^2\cos 2\theta_1 +a\omega_1\sin 2\theta_1 \right)+\frac{1-\cos 2\theta_1}{2}+\alpha\beta\sin\varphi\left[\omega_1\sin 2\theta_1+a(1-\cos 2\theta_1)\right]}{\beta^2\left(\omega_1^2\cos 2\theta_1-a\omega_1\sin 2\theta_1\right)+\frac{1-\cos 2\theta_1}{2}-\alpha\beta\sin\varphi\left[\omega_1\sin 2\theta_1-a(1-\cos 2\theta_1)\right]}.\label{Eq:PTm2}
\end{equation}
After inserting Eq.~(\ref{Eq:PTm2}) into Eq.~(\ref{Eq:dcdm}), we obtain
\begin{equation}
 \tan 2\theta_1= -\frac{\left(\alpha^2-\beta^2\right)\omega_1}{a+2\alpha\beta\sin\varphi}.\label{Eq:tanm}
\end{equation}
Note that the period of $\tan 2\theta_1$ is $\frac{\pi}{2}$ with respect to $\theta_1$, while the period of $C_{l_1} \left(|\phi(t)\rangle\right)$ is $T_{\mathcal{PT}}$=$\pi/(\omega_1 s)$ with respect to $t$. Because of $\theta_1$=$\omega_1 st$, the period $T_{\mathcal{PT}}$=$\pi/(\omega_1s)$ can be expressed as $T_{\theta_1}$=$\pi$ with respect to $\theta_1$. Thus, one period of $C_{l_1} \left(|\phi(t)\rangle\right)$ includes two periods of $\tan 2\theta_1$; that is, there are two different values of $\theta_1$ (or $t$) satisfying Eq.~(\ref{Eq:tanm}) or Eq.~(\ref{Eq:PTdCdm}) within one period of $C_{l_1} \left(|\phi(t)\rangle\right)$.

Now, we consider the case of $\frac{dm_1}{d\theta_1}=0$. Based on $m_1=x_1/y_1$, one has
\begin{equation}
 \frac{dm_1}{d\theta_1}=\frac{x_1^{\prime}y_1-x_1 y_1^{\prime}}{y_1^2},\label{Eq:dmdthexy}
\end{equation}
where $x_1^{\prime}=\frac{dx_1}{d\theta_1}$ and $y_1^{\prime}=\frac{dy_1}{d\theta_1}$. It follows from Eq.~(\ref{Eq:PTCxy1}) that:
\begin{eqnarray}
 x_1^{\prime}=\frac{1}{\omega_1^2}\left[\alpha^2\left(-2\omega_1^2\sin 2\theta_1+2a\omega_1\cos 2\theta_1\right)+\sin 2\theta_1 +\alpha\beta\sin\varphi\left(2\omega_1\cos 2\theta_1+2a\sin 2\theta_1\right)\right],\nonumber\\
 y_1^{\prime}=\frac{1}{\omega_1^2}\left[\beta^2\left(-2\omega_1^2\sin 2\theta_1-2a\omega_1\cos 2\theta_1\right)+\sin 2\theta_1 -\alpha\beta\sin\varphi\left(2\omega_1\cos 2\theta_1-2a\sin 2\theta_1)\right)\right].\label{Eq:dxdy}
\end{eqnarray}
 Substituting Eq.~(\ref{Eq:PTCxy1}) and Eq.~(\ref{Eq:dxdy}) into Eq.~(\ref{Eq:dmdthexy}), one can easily find that the condition for $dm_1/d\theta_1=0$ is:

\begin{eqnarray}
 \lefteqn{\left[2a\omega_1^2\alpha^2\beta^2-a(4\alpha^2\beta^2\sin^2\varphi+1)-\alpha\beta\sin\varphi(3a^2+1)\right]\tan^2\theta_1 -\omega_1 (1-2\beta^2)(1-2a\alpha\beta\sin\varphi)\tan \theta_1}\hspace*{440pt} \nonumber\\
 \lefteqn{+2a\omega_1\alpha^2\beta^2+\alpha\beta\sin\varphi=0.}\hspace*{440pt}\label{Eq:m1}
\end{eqnarray}

Thus, the discriminant of Eq.~(\ref{Eq:m1}) is given by:
\begin{equation}
 \Delta=g+4(p+q)r, \label{Eq:Delta}
\end{equation}
with
\begin{eqnarray}
 \lefteqn{g=\omega_1^2 (\alpha^2-\beta^2)^2(1-2a\alpha\beta\sin\varphi)^2,}\hspace*{200pt} \nonumber\\
  \lefteqn{p=4a\alpha^2\beta^2\sin^2\varphi+\alpha\beta\sin\varphi(3a^2+1),}\hspace*{200pt} \nonumber\\
  \lefteqn{q=a\left[(\alpha^2-\beta^2)^2+2(1+a^2)\alpha^2\beta^2\right],}\hspace*{200pt} \nonumber\\
  \lefteqn{r=2a\omega_1^2\alpha^2\beta^2+\alpha\beta\sin\varphi.}\hspace*{200pt}
\label{Eq:Delta1}
\end{eqnarray}
 Here, $g \geq 0$ and $q > 0$. Without loss of generality, we consider $\sin \varphi \geq 0$ and $\alpha, \beta \in \left(0,~1\right)$. In this case, $p \geq 0$ and $r > 0$. Hence, we have $\Delta > 0$, which implies that $\tan \theta_1$ has two different values to satisfy either Eq.~(\ref{Eq:m1}) or Eq.~(\ref{Eq:PTdmdthe0}). As mentioned above, the period $T_{\mathcal{PT}}$=$\frac{\pi}{s\sqrt{1-a^2}}$ of $C_{l_1}\left(|\phi(t)\rangle\right)$ can be expressed as $T_{\theta_1}$=$\pi$ with respect to $\theta_1$. Note that the period of $\tan \theta_1$ and the period of  $C_{l_1}\left(|\phi(t)\rangle\right)$ are $\pi$ with respect to $\theta_1$, and $\tan \theta_1$ has two different values to satisfy Eq.~(\ref{Eq:PTdmdthe0}). Thus, there exist two different values $\theta_1$ (or $t$) to satisfy Eq.~(\ref{Eq:PTdmdthe0}) within one period of $C_{l_1}\left(|\phi(t)\rangle\right)$.

From the above discussion, one can conclude that for a wide rangle of initial states $\alpha|H\rangle +\beta e^{i\varphi} |V\rangle$, with $\alpha, \beta \in \left(0,~1\right)$ and $\sin \varphi \geq 0$, the $ \frac{dC_{l_1} \left(|\phi(t)\rangle\right)}{dt}$ has four zero points in one period (i.e., $T$=$\frac{\pi}{s\sqrt{1-a^2}}$) of coherent evolution. Therefore, in the $\mathcal{PT}$-symmetric-unbroken regime, there indeed exists the phenomenon of two backflows of coherence in a period of coherent evolution (e.g., see Supplementary Figure~\ref{afig1}).\\
\begin{figure}[!htbp]
\setlength{\belowcaptionskip}{-0.15cm}
       \includegraphics[width=0.4\linewidth]{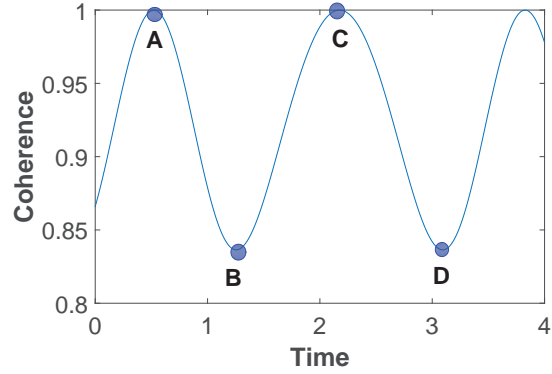}
\caption{The points $A$, $B$, $C$ and $D$ are four extreme points within one period. Note that in the $\mathcal{PT}$-symmetric-unbroken regime, there are two backflows of coherence inside a simple period of coherent evolution.}
\label{afig1}
\end{figure}

 \begin{center}
 \makeatletter
 \renewcommand{\section}{\textbf{Supplementary Note 4:~}}
\section{\textbf{ Proof for the characteristics of backflow in the $\mathcal{APT}$-symmetric-unbroken regime}}\\
 \end{center}

 \indent For an arbitrary initial state $|\phi\rangle=\alpha|H\rangle+\beta e^{i\varphi}|V\rangle$, the coherence of the evolved state in the $\mathcal{APT}$-symmetric systems is given by Eq.~(\ref{Eq:CAPT1}). In the $\mathcal{APT}$-symmetric-unbroken regime (i.e., $a>1$), $A$, $B$ and $C$  are given by Eq.~(\ref{Eq:APTABC}). In view of Eq.~(\ref{Eq:CAPT2}), the derivative of $ C_{l_1} \left(|\phi(t)\rangle\right)$ can be expressed as:
\begin{equation}
 \frac{dC_{l_1} \left(|\phi(t)\rangle\right)}{dt}=\frac{dC_{l_1} \left(|\phi(t)\rangle\right)}{dm_3}\times \frac{dm_3}{d\theta_3}\times \frac{d\theta_3}{dt}.\label{Eq:aptdcdt1}
\end{equation}
Note that $\frac{d\theta_3 }{dt}=\omega_3 s>0$. Thus, to meet $\frac{dC_{l_1} \left(|\phi(t)\rangle\right)}{dt}=0$, it follows from Eq.~(\ref{Eq:aptdcdt1}):
\begin{equation}
\frac{dC_{l_1} \left(|\phi(t)\rangle\right)}{dm_3}=0,
\end{equation}
or
\begin{equation}
\frac{dm_3}{d\theta_3}=0.\label{Eq:admdthe1}
\end{equation}

First, we consider the case when $\frac{dC_{l_1} \left(|\phi(t)\rangle\right)}{dm_3}=0$. According to Eq.~(\ref{Eq:CAPT2}), we have
\begin{equation}
 \frac{dC_{l_1} \left(|\phi(t)\rangle\right)}{dm_3}=\frac{1-m_3}{(1+m_3)^2\sqrt{m_3}}=0.\label{Eq:adcdm}
\end{equation}
Because of $m_3=x_3/y_3$ and according to Eq.~(\ref{Eq:APTxy1}), we have
\begin{equation}
 m_3=\frac{\left[\omega_3^2\alpha^2+\alpha\beta\omega_3\cos\varphi\sin 2\theta_3+\frac{1-\cos 2\theta_3}{2}\left(1-a\alpha\beta\sin\varphi\right)\right]}{\left[\omega_3^2\beta^2+\alpha\beta\omega_3\cos\varphi\sin 2\theta_3+\frac{1-\cos 2\theta_3}{2}\left(1-a\alpha\beta\sin\varphi\right)\right]}.\label{Eq:APTm}
\end{equation}
Substituting Eq.~(\ref{Eq:APTm}) into Eq.~(\ref{Eq:adcdm}) leads to
\begin{equation}
 \alpha^2-\beta^2=0.\label{Eq:aptm1}
\end{equation}
 In general, Eq.~(\ref{Eq:aptm1}) is not satisfied for an arbitrary initial state $\alpha|H\rangle+\beta e^{i\varphi}|V\rangle$.

 Now, we consider the other case of $dm_3/d\theta_3=0$. Because of $m_3=x_3/y_3$ and according to Eq.~(\ref{Eq:APTxy1}), one has
\begin{equation}
\frac{dm_3}{d\theta_3}=\frac{x_3^\prime y_3- x_3 y_3^\prime}{y_3^2},\label{Eq:admdthe}
\end{equation}
where
\begin{eqnarray}
 x_3^\prime=\frac{1}{\omega_3^2}\left[2\alpha\beta\omega_3\cos\varphi \cos 2\theta_3 +\sin 2\theta_3 \left(1-a\alpha\beta\sin\varphi\right)\right],\nonumber\\
 y_3^\prime=\frac{1}{\omega_3^2}\left[2\alpha\beta\omega_3\cos\varphi \cos 2\theta_3 +\sin 2\theta_3\left(1-a\alpha\beta\sin\varphi\right)\right].\label{Eq:adxdy}
\end{eqnarray}
According to Eqs.~(\ref{Eq:APTxy1},~\ref{Eq:admdthe},~\ref{Eq:adxdy}), one can easily find that the condition for $dm_1/d\theta_1=0$ is:
\begin{equation}
 \tan 2\theta_3=-\frac{2\alpha\beta\omega_3\cos\varphi}{1-a\alpha\beta\sin\varphi}.\label{Eq:atanm}
\end{equation}
Because the period of $\tan 2\theta_3$ is $\frac{\pi}{2}$ and the period of $C_{l_1} \left(|\phi(t)\rangle\right)$ is $T_{\theta_3}$=$\pi$ (i.e., $T_{\mathcal{APT}}$=$\frac{\pi}{s\sqrt{a^2-1}}$), one period of $C_{l_1} \left(|\phi(t)\rangle\right)$ includes two periods of $\tan 2\theta_3$. Thus, there exist two different values of $\theta_3$ (or $t$) satisfying Eq.~(\ref{Eq:atanm}) or Eq.~(\ref{Eq:admdthe1}) within one period of $C_{l_1} \left(|\phi(t)\rangle\right)$.
\begin{figure}[!htbp]
       \includegraphics[width=0.4\linewidth]{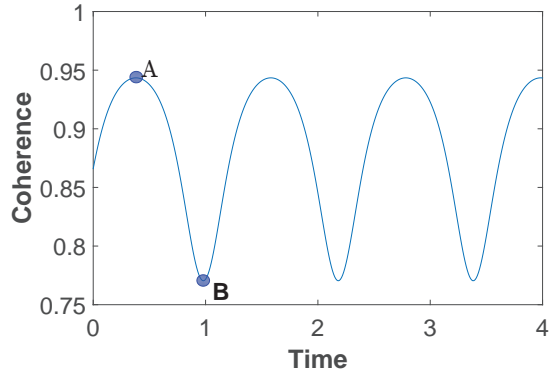}
\caption{The points $A$ and $B$ are two extreme points in one period. The coherent oscillation of quantum states in the $\mathcal{APT}$-symmetric-unbroken regime has only one backflow within one period.}
\label{afig2}
\end{figure}

 From the above discussion, one can conclude that $ \frac{dC_{l_1} \left(|\phi(t)\rangle\right)}{dt}$ has two zero points in one period (i.e., $T_{\mathcal{APT}}$=$\frac{\pi}{s\sqrt{a^2-1}}$) of coherent evolution. Therefore, the coherent oscillation of quantum states in the $\mathcal{APT}$-symmetric-unbroken regime has only one backflow within one period (eg., see Supplementary Figure~\ref{afig2}).

\clearpage

\begin{center}
 \makeatletter
 \renewcommand{\section}{\textbf{Supplementary Note 5:~}}
\section{\textbf{ Experimental implementation of the loss operator $L$}}
 \end{center}

\indent As illustrated in Supplementary Figure \ref{afig3}, we experimentally implement the loss operator $L$ by a combination of two beam displacers ($\rm BD_1$ and $\rm BD_2$) and two half-wave plates ($\rm HWP_1$ and $\rm HWP_2$). Here, the optical axes of the BDs are cut so that the vertically polarized photons are transmitted directly, while the horizontally polarized photons are displaced into the lower path. In addition, the $\rm HWP_1$ and $\rm HWP_2$ with setting angles $\xi_i$ and $\xi_j$ are, respectively, inserted into the upper and lower paths between the two BDs. The rotation operations on the photon polarization states, performed by the $\rm HWP_1$ and $\rm HWP_2$, are given as follows:
\begin{figure}[!htbp]
 \includegraphics[width=0.55\linewidth]{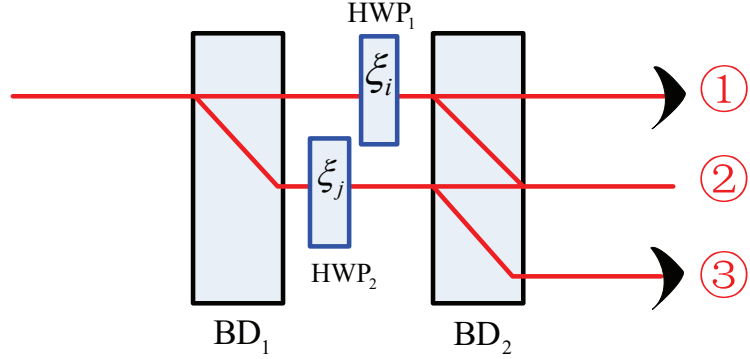}
\caption{Experimental setup to realize a loss operator, where $\xi_i$ and $\xi_j$ are the two tunable setting angles for the half-wave plates $\rm HWP_1$ and $\rm HWP_2$, respectively.}
\label{afig3}
\end{figure}
\begin{equation}
R_{\rm HWP}\left(\xi_i\right)=\left(
                             \begin{array}{cc}
                               \cos 2\xi_i & \sin 2\xi_i \\
                               \sin 2\xi_i & -\cos 2\xi_i \\
                             \end{array}
                           \right),~~~~
 R_{\rm HWP}\left(\xi_j\right)=\left(
                             \begin{array}{cc}
                               \cos 2\xi_j & \sin 2\xi_j \\
                               \sin 2\xi_j & -\cos 2\xi_j \\
                             \end{array}
                           \right).
\end{equation}
In this case, when a horizontally polarized photon passes through the experimental setup, one can find that
\begin{equation}
|H\rangle~\xrightarrow{\rm BD_1}~|H\rangle_{\rm lower}~\xrightarrow{R_{\rm HWP}(\xi_j)}~R_{\rm HWP}(\xi_j)|H\rangle ~\xrightarrow{\rm BD_2}~\cos 2\xi_j|H\rangle_3 +\sin 2\xi_j |V\rangle_2,\label{BDH}
\end{equation}
where the subscript ``lower'' represents the lower path between the two BDs, while subscripts ``2'' and ``3'' represent the two paths $2$ and $3$ after the second BD, respectively. Similarly, when a vertically polarized photon pass the experimental setup, one can find that
\begin{equation}
|V\rangle ~ \xrightarrow{\rm BD_1} ~ |H\rangle_{\rm upper} ~ \xrightarrow{R_{\rm HWP}(\xi_i)}~ R_{\rm HWP}(\xi_i)|V\rangle ~ \xrightarrow{\rm BD_2} ~ \sin 2\xi_i|H\rangle_2 -\cos 2\xi_i |V\rangle_1,\label{BDV}
\end{equation}
 where the subscript ``upper'' represents the upper path between the two BDs, while subscripts ``1'' and ``2'' represent the two paths $1$ and $2$ after the second BD, respectively. That is, only horizontally polarized photons in the upper path and vertically polarized photons in the lower path are transmitted through the second BD and then combined onto path $2$, while the other photons transmitted onto path $1$ or $3$ are blocked, i.e., they are discarded and lost from the system.

 In this sense, according to Eqs. (\ref{BDH}) and (\ref{BDV}), when the input photon is  initially in the state $|\phi\rangle_{\rm in}=\alpha|H\rangle +\beta e^{i\varphi}|V\rangle$, then the output photon appearing in the path $2$ would be in the state $|\phi\rangle_{\rm out} = \alpha \sin 2\xi_j |V\rangle_2 + \beta e^{i\varphi} \sin 2\xi_i |H\rangle_2$. It is obvious that this state transformation can be written as $|\phi\rangle_{\rm out}=L|\phi\rangle_{\rm in}$, with a polarization-dependent photon loss operator $L$, given by
\begin{equation}
L\left(\xi_i,~\xi_j\right)=\left(
                             \begin{array}{cc}
                               0 & \sin 2\xi_i \\
                               \sin 2\xi_j & 0 \\
                             \end{array}
                           \right),
\end{equation}
where $\xi_i$ and $\xi_j$ are, respectively, the two tunable setting angles for the half-wave plates $\rm HWP_1$ and $\rm HWP_2$ (Supplementary Figure~\ref{afig3}).\\
\clearpage
 \begin{center}
 \makeatletter
 \renewcommand{\section}{\textbf{Supplementary Note 6:~}}
\section{\textbf{ Coherence flow for two-qubit $\mathcal{PT}$- and anti-$\mathcal{PT}$- symmetric systems}}
 \end{center}

We have numerically simulated the dynamics of coherence for two-qubit $\mathcal{PT}$/$\mathcal{APT}$ systems. As shown in Supplementary Figures \ref{afig4}(a, c), there exist different periodic oscillations of coherence (including one coherence backflow, two coherence backflows, and multiple coherence backflows in one period) for $\mathcal{PT}$/$\mathcal{APT}$-symmetric systems in the unbroken regime. In addition, as illustrated in Supplementary Figures \ref{afig4}(b, d), there exists PSV for both $\mathcal{PT}$ -and $\mathcal{APT}$ -symmetric systems in the broken regime, which are independent of the initial states.

\begin{figure}[!htbp]
\begin{flushleft}\hspace*{40pt}
\setlength{\belowcaptionskip}{-0.7cm}
    \subfigure{
       \includegraphics[scale=0.43]{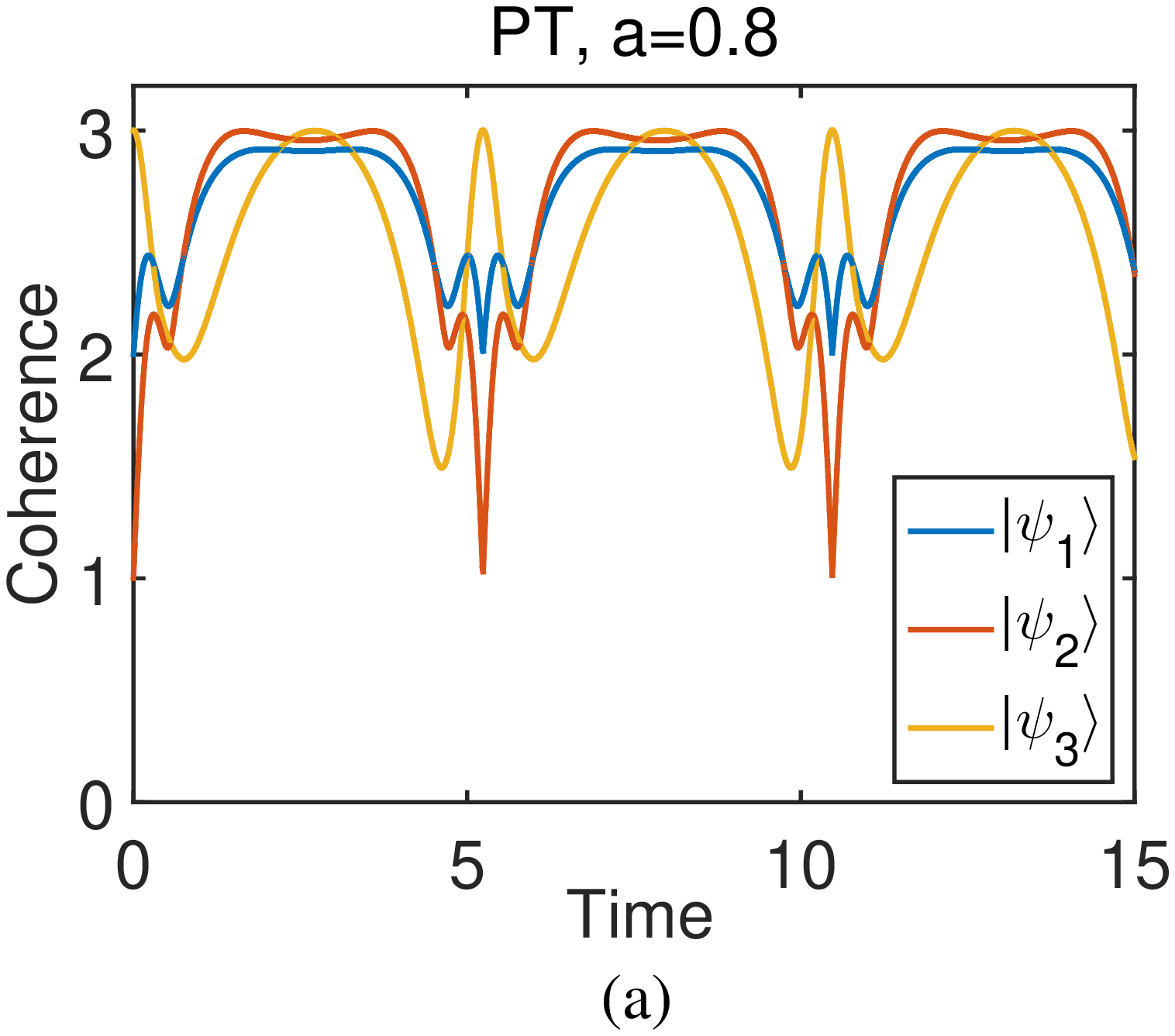}}\hspace{10mm}
    \label{1a}
    \subfigure{
        \includegraphics[scale=0.43]{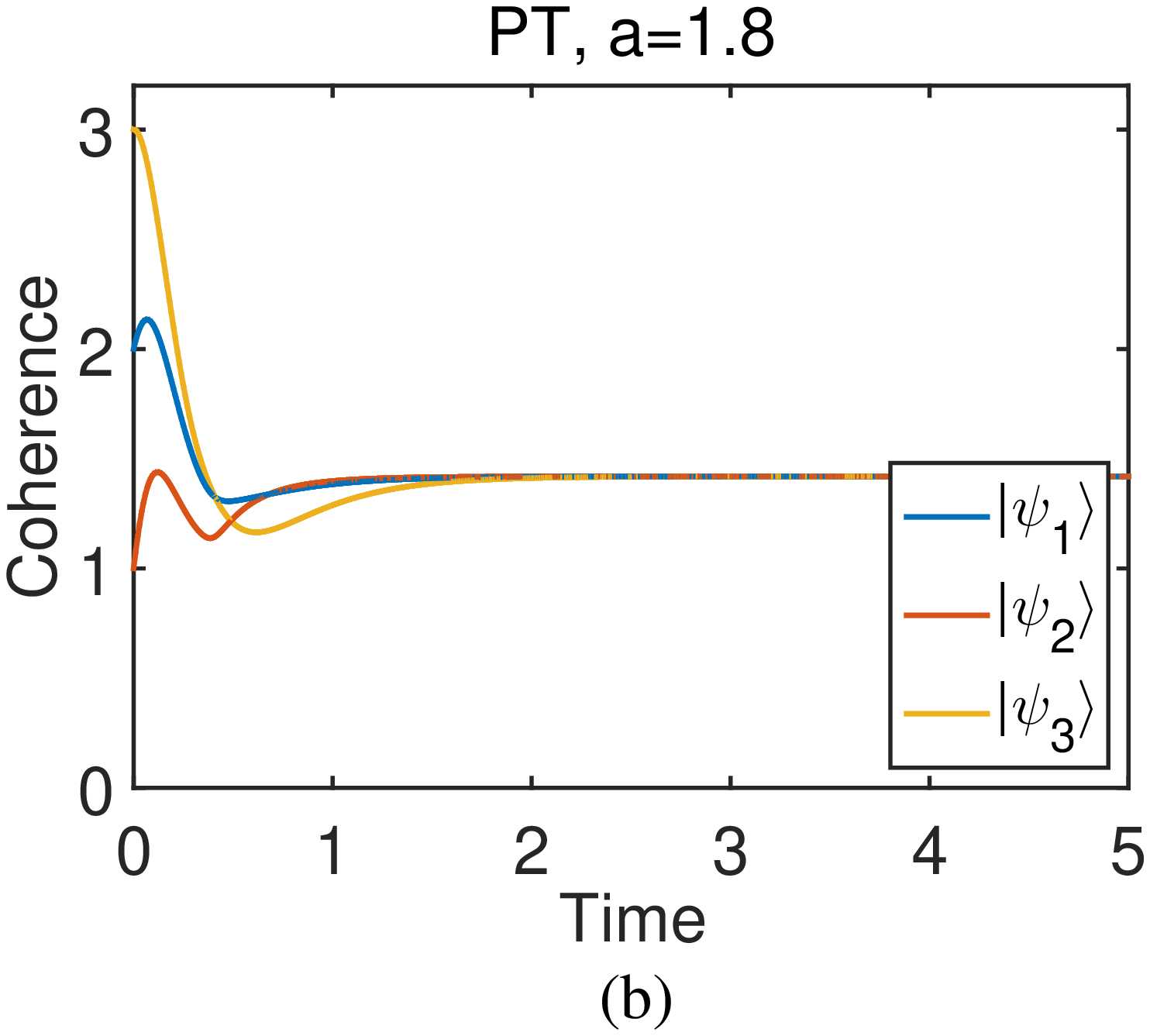}}\\
    \end{flushleft}
    \begin{flushleft}\hspace*{40pt}
    \subfigure{
        \includegraphics[scale=0.43]{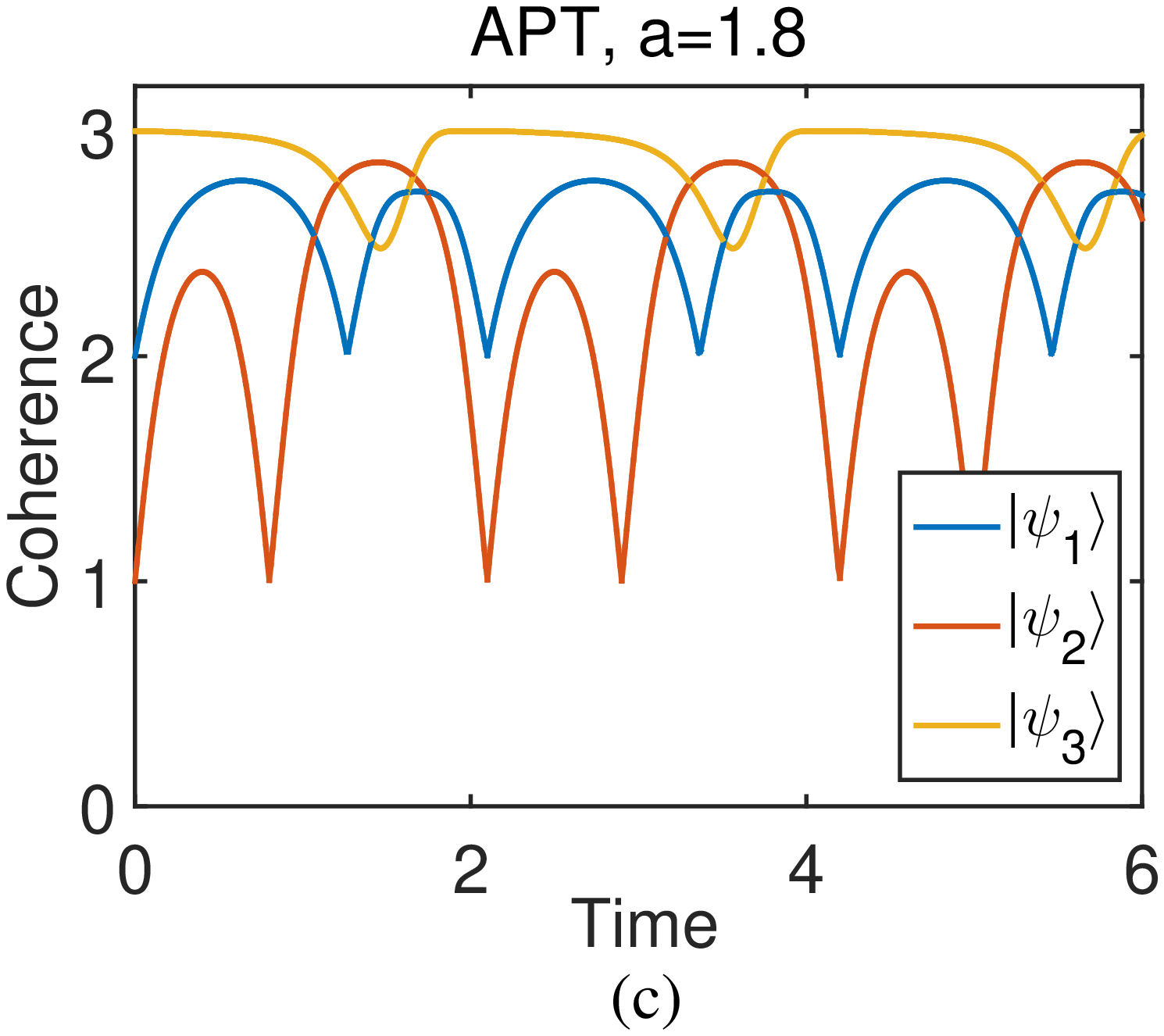}}\hspace{12mm}
    \subfigure{
        \includegraphics[scale=0.43]{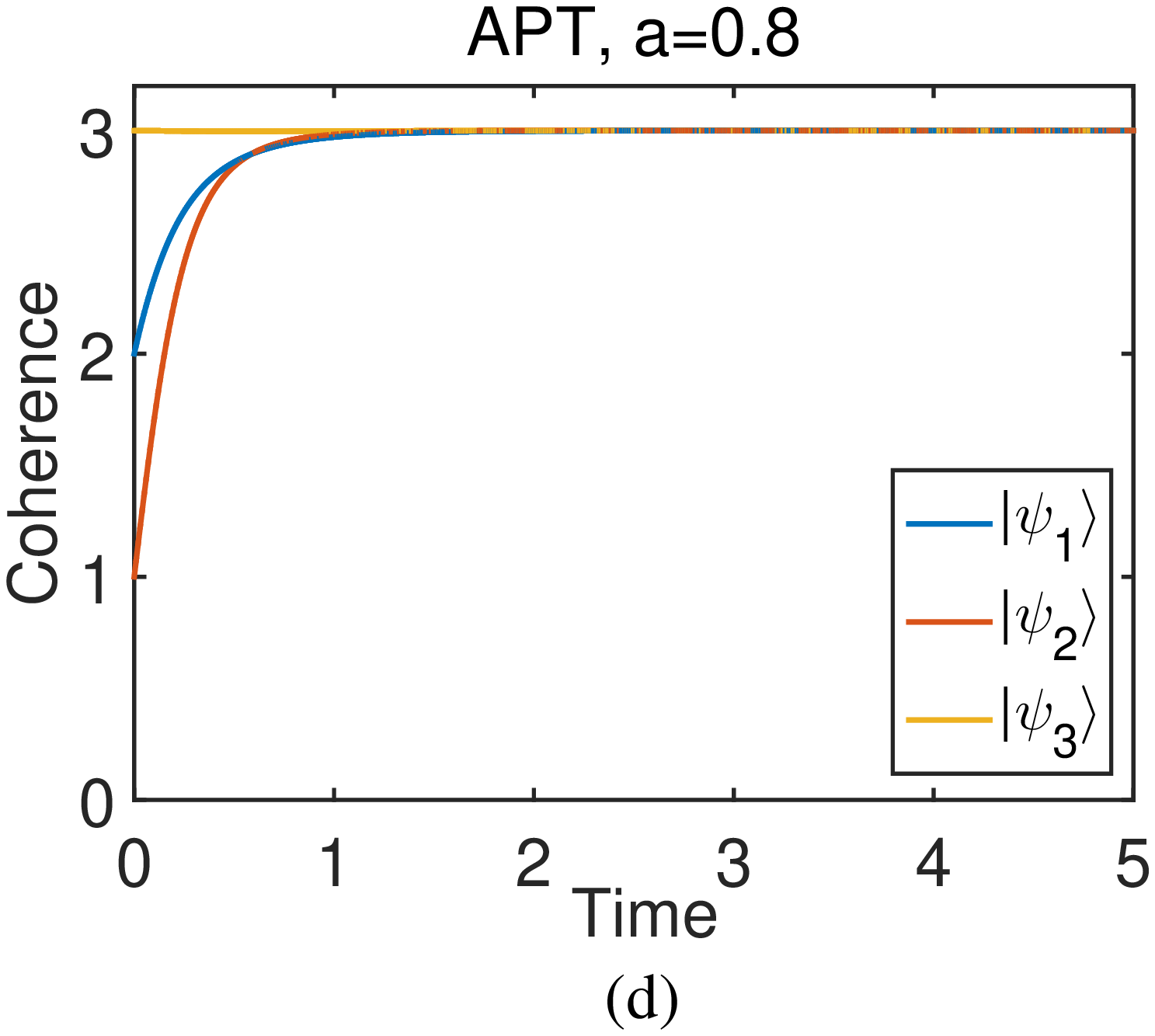}}
\caption{The evolution of coherence for three different initial states in a two-qubit $P\mathcal{T}$ /$\mathcal{APT}$ -symmetric system. We consider the two qubits undergoing the same $\mathcal{PT}$/$\mathcal{APT}$ -symmetric dynamic process, i.e., the parameters $a$ involved in the Hamiltonians (1) and (2) of the main text are set to be the same for both qubits. (a) $a=0.8$, the $\mathcal{PT}$ symmetry unbroken regime; (b) $a=1.8$, the $\mathcal{PT}$ symmetry broken regime; (c) $a=1.8$, the $\mathcal{APT}$ symmetry unbroken regime; (d) $a=0.8$, the $\mathcal{APT}$ symmetry broken regime. The three initial states are $|\psi_1\rangle=\frac{1}{\sqrt{3}}\left(|00\rangle+|01\rangle + |11\rangle\right)$ (blue curves), $|\psi_2\rangle=\frac{1}{\sqrt{2}}\left(|00\rangle + e^{i\pi/5}|11\rangle\right)$ (red curves), and $|\psi_3\rangle=\frac{1}{2}\left(|00\rangle +|01\rangle +|10\rangle +  e^{i\pi/5} |11\rangle\right)$ (yellow curves).}\label{afig4}
\end{flushleft}
\end{figure}
\end{widetext}

\end{document}